\documentclass[iop]{emulateapj}

\usepackage{amssymb} 
\usepackage{amsmath} 
\usepackage{graphicx}
\usepackage{latexsym} 
\usepackage{dcolumn}
\usepackage{bm}
\input{epsf} 

\newcommand{\be}{\begin{equation}}
\newcommand{\ee}{\end{equation}} 
\newcommand{\ba}{\begin{eqnarray}}
\newcommand{\ea}{\end{eqnarray}} 

\newcommand{\bfi}{\begin{figure} \epsfxsize=8cm \epsffile}
\newcommand{\bfig}{\begin{figure*} \epsfxsize=15cm \epsffile}
\newcommand{\efi}{\end{figure}} 
\newcommand{\efig}{\end{figure*}}
\newcommand{\bi}{\begin{itemize}} 
\newcommand{\ei}{\end{itemize}}
\newcommand{\mpch}{h^{-1} {\rm Mpc}} 
\newcommand{\hmpc}{h {\rm Mpc}^{-1}}

\newcommand{\Msun}{M_{\odot}}
\newcommand{\dif}{\mathrm{d}} 
\newcommand{\etal}{{\it et al.}}

\newcommand{\vecx}{{\bm{x}}}
\newcommand{\vecxi}{{\bm{\xi}}}

\slugcomment{submitted to ApJ}

\makeatletter

\shorttitle{Halo Nonlinear Reconstruction}
\shortauthors{Yu et al.}

\begin{document}

\title{Halo Nonlinear Reconstruction}

\author{Yu Yu\altaffilmark{1,*}}
\altaffiltext{1}{Key Laboratory for Research in Galaxies and Cosmology,
Shanghai Astronomical Observatory, Chinese Academy of Sciences, 80
Nandan Road, Shanghai, 200030, China}
\altaffiltext{*}{yuyu22@shao.ac.cn}
\author{Hong-Ming Zhu\altaffilmark{2,3}}
\altaffiltext{2}{Key Laboratory for Computational Astrophysics, National Astronomical Observatories,
Chinese Academy of Sciences, 20A Datun Road, Beijing 100012, China}
\altaffiltext{3}{University of Chinese Academy of Sciences, Beijing 100049, China}
\author{Ue-Li Pen\altaffilmark{4,5,6,7}}
\altaffiltext{4}{Canadian Institute for Theoretical Astrophysics, University of Toronto,
60 St. George Street, Toronto, Ontario M5S 3H8, Canada}
\altaffiltext{5}{Dunlap Institute for Astronomy and Astrophysics, University of Toronto,
50 St. George Street, Toronto, Ontario M5S 3H4, Canada}
\altaffiltext{6}{Canadian Institute for Advanced Research, CIFAR Program in
Gravitation and Cosmology, Toronto, Ontario M5G 1Z8, Canada}
\altaffiltext{7}{Perimeter Institute for Theoretical Physics, 31 Caroline Street North, Waterloo, Ontario, N2L 2Y5, Canada}

\begin{abstract}
We apply the nonlinear reconstruction method (Zhu \etal, arXiv:1611.09638) to simulated halo fields.  For halo number density $2.77\times 10^{-2}$ $(\mpch)^{-3}$ at $z=0$, corresponding to the SDSS main sample density, we find that the scale where the noise saturates the linear signal is improved to $k\gtrsim0.36\ \hmpc$, which is a factor of $2.29$ improvement in scale, or $12$ in number of linear modes.  The improvement is less for higher redshift or lower halo density. We expect this to substantially improve the BAO accuracy of dense, low-redshift surveys, including the SDSS main sample, 6dFGS and 21cm intensity mapping initiatives.
\end{abstract}



\section{Introduction}
\label{sec:introduction}

Measuring the expansion history and structure growth has been a major focus of modern observational cosmology.
A lot of information about our universe is being decoded from the current surveys, like BOSS 
(\cite{BOSScomplete}),
6dFGS (\cite{6dFGS-DR3}), and etc.
Ambitious on-going and future surveys, like eBOSS\footnote{eBOSS, http://www.sdss.org/surveys/eboss/}, DESI 
(\cite{DESIsci}), PFS 
(\cite{PFScosmology}), CHIME
(\cite{Bandura14}), 
HIRAX (\cite{HIRAX16}),
BINGO (\cite{BINGO16}),
Tianlai\footnote{Tianlai, http://tianlai.bao.ac.cn},
HETDEX\footnote{HETDEX, http://www.hetdex.org/}, Euclid\footnote{EUCLID, http://sci.esa.int/euclid/}, WFIRST\footnote{WFIRST, http://wfirst.gsfc.nasa.gov/}, etc,
 will further extend the exploration in both width and depth.
These surveys are expected to bring us significant improvement in the understanding of our universe.

Most of the future surveys are aiming at the high-redshift universe.
These surveys will observe matter distributions with less nonlinear evolution in a huge cosmic volume.
However, only very bright objects/features could be detected in these surveys.
On the contrary, the low-redshift universe is relatively easy to observe, but suffers from significant nonlinear effects and limited volume.
The late-time nonlinear evolution is a complicated process and the statistics are difficult to model.
This induces systematics in the statistics of cosmic probes (e.g., the broadening and shifting of the baryon acoustic oscillation (BAO) peak in the galaxy two-point correlation function).
It also prevents a simple mapping from the final state to the initial conditions that are predicted by theories.
As a result, a large portion of the cosmic information is encoded into complicated high-order statistics, and even worse, some of it is lost (e.g., \cite{Rimes05,Rimes06,Neyrinck06,Carron12a}).

Part of the information loss could be recovered by a process known as ``reconstruction'' (e.g., \citet{Eisenstein07}).
The nonlinear density field is smoothed on the linear scale ($\sim 10\ \mpch$) to make the Zel'dovich approximation valid.
The linear displacement is estimated and used to move the galaxies back and to move a random sample in the same way.
These two new fields form the reconstructed density field,
which has a sharper BAO peak, leading to more stringent cosmological parameter constraints.
We call the above reconstruction method as ``the standard reconstruction method''.
In the literature, this standard reconstruction method was theoretically understood and modeled (e.g., \cite{Noh09}, \cite{Padmanabhan09},  \cite{White15}, \cite{Seo16}), tested against numerical simulations (e.g., \cite{Seo08}, \cite{Seo10}, \cite{Mehta11}, \cite{Achitouv15}, \cite{Schmittfull15}, \cite{Obuljen16}), and applied on observations (e.g., {\cite{Padmanabhan12}, \cite{xuxy13}, \cite{Anderson14}, \cite{Kazin14}, \cite{Ross15}, \cite{Beutler16,Beutler17}, \cite{Hinton17}) extensively.

\citet{zhuhm16c} presented a direct approach to nonparametrically reconstruct the linear density field,
by solving for a unique displacement potential consistent with the nonlinear density map and positive definite coordinate transformation.
Different from \citet{Eisenstein07}, the reconstructed (nonlinear) displacement is not used to move the galaxy positions.
Instead, the reconstructed density field is directly derived by the negative divergence of the reconstructed displacement.
Hereafter, we call this reconstruction process as {\it nonlinear reconstruction}.
Using the simulated dark matter density field, \citet{zhuhm16c} found that the reconstructed density field recovers the coherence with linear initial conditions up to $k\sim 1\ \hmpc$, a factor of $\sim 6$ improvement in scale compared to the nonlinear density field.
The information content is also found to be increased by a factor of $\sim 50$ after the nonlinear reconstruction (see \citet{panqy17}).
\cite{yuhr17} quantified the limits of all Lagrangian reconstruction methods by measuring the correlations in Lagrangian space (i.e., the correlations between the initial displacement and final displacement) in simulations.

Nonlinear reconstruction generalizes the application of linear displacement theory to fully nonlinear fields, potentially substantially expanding the BAO and redshift space distortion (RSD) information content of dense large-scale structure surveys.
As shown in \citet{zhuhm16c}, the noise part of the reconstructed field dominates over the linear signal at $k\gtrsim0.6\ \hmpc$, suggesting that all BAO peaks may be recovered from the present day density field, opening up the potential of recovering cosmic information including BAO down to the Poisson noise limit.
To apply this novel reconstruction method to observations, we need to consider the reconstruction from galaxy/halo density fields. 
This paper presents the performance test on the simulated halo fields with three number densities and three redshifts.

This article is organized as follows.
In Sect. \ref{sec:recon}, we briefly introduce the reconstruction method.
The performance tests on various situations are presented in Sect. \ref{sec:result}.
We summarize the result and discuss future directions in Sect. \ref{sec:conclusion}.


\section{Reconstruction Algorithm}
\label{sec:recon}


The basic idea of the reconstruction is to build a bijective curvilinear coordinate system $\vecxi\equiv(\xi_1,\xi_2,
\xi_3)$, where the mass per volume element is constant,
\be
\rho(\vecxi)\dif^3\vecxi={\rm constant}\ .
\ee
We call this curvilinear coordinate system as {\it potential isobaric gauge/coordinates}.
It becomes analogous to ``synchronous gauge'' and ``Lagrangian coordinates'' before shell crossing, but allows a unique mapping even after shell crossing.
In this scenario, the mass element at final physical position $\vecx$ comes from the estimated initial Lagrangian position given by the potential isobaric coordinate $\vecxi$.
In the following, we use Latin indices to denote Cartesian coordinate labels $x^i$ and Greek indices to denote the curvilinear coordinates $\xi^\mu$.

Since we attempt to follow the potential flow instead of the vorticity, we define a coordinate transformation that is a pure gradient,
\be
x^i=\xi^\mu\delta_\mu^i+\Delta x^i\ ,
\ee
where the displacement
\be
\Delta x^i\equiv\frac{\partial\phi}{\partial\xi^\nu}\delta^{i\nu}\ .
\ee
We call $\phi$ as the displacement potential.

Since the displacement from the initial Lagrangian position to the final Eulerian position can be large,
it is difficult to obtain the solution directly.
One efficient and robust algorithm is the moving mesh approach, which is originally introduced for the adaptive particle-mesh $N$-body algorithm and the moving mesh hydrodynamics algorithm (see \cite{Pen95,Pen98}).
These algorithms attempt to evolve the curvilinear coordinate along with the matter/energy density field to maintain constant mass/energy-resolution.
In our case, this approach solves for the displacement potential perturbatively and iteratively.
The evolution of the curvilinear coordinate system is determined by a linear elliptic evolution equation
\be
\partial_\mu(\rho\sqrt{g}e^\mu_i\delta^{i\nu}\partial_\nu\dot\phi)=\Delta\rho\ ,
\label{eqn:elliptic}
\ee
where $\sqrt{g}=\det(e^i_\mu)$ is the Jabobian of the transformation matrix $e^i_\mu=\partial x^i/\partial\xi^\mu$ and $\Delta\rho=\bar\rho-\rho\sqrt{g}$.
The time derivative here is relative to the iteration step.
We obtain the change of the displacement potential $\Delta\phi=\dot\phi\Delta t$ at each iteration step and then update the density field in the new coordinate frame.
The final solution is given by 
\be
\phi=\Delta\phi^{(1)}+\Delta\phi^{(2)}+\Delta\phi^{(3)}+\cdots\ ,
\ee
where $\Delta\phi^{(i)}$ is the result from the $i$th iteration.
We also implement the smoothing and limiting schemes to guarantee the triad $e^\mu_i$ is positive definite at each step.
Note that different from the smoothing kernel used in the standard reconstruction method, this smoothing is to keep the algorithm stable at each step and only influences the efficiency of the algorithm but not the final result of the reconstruction.
The elliptic equation can be solved using the multigrid algorithm described also in \cite{Pen95}.

The above process ensures that the coordinate transform is positive definite.
The coordinate lines will not cross and the eigenvalues for this coordinate transform are always positive.
For the reconstruction from the dark matter density field and in the case that dark matter particles follow a irrotational potential flow and no shell crossing happens,
 the reconstructed displacement is exact up to a global spatial translation.
However, shell crossing happens in the nonlinear regime.  This reconstruction algorithm gives an approximate solution to the true displacement.

We are aiming at reconstructing a density field with more linear information.
We define the negative Laplacian of the reconstructed displacement potential as the {\it reconstructed density field},
\be
\label{eqn:recon}
\delta_r(\vecxi)\equiv-\nabla_\vecxi\cdot\Delta\vecx(\vecxi)=-\nabla_\vecxi^2\phi\ .
\ee
See \cite{zhuhm16c} for a more detailed physical interpretation of this reconstruction process and the relation with \cite{Eisenstein07}.



\section{Implementation and Result}
\label{sec:result}

\subsection{Halo density field}
\label{sec:massassign}

\begin{table*}
\caption{Detailed Information for Various Halo Samples Used in the Performance Test}
\begin{tabular}{p{1.6cm}p{4.1cm}<{\centering}p{4.1cm}<{\centering}p{4.1cm}<{\centering}p{2.3cm}<{\centering}}
\hline\hline
$M_{\rm min}$, $b$ & $n_h=2.77\times 10^{-2} (\mpch)^{-3}$ & $n_h=2.77\times 10^{-3} (\mpch)^{-3}$ & $n_h=2.77\times 10^{-4} (\mpch)^{-3}$ & $M_{\rm max}$\\
\hline
$z=0.0$    & $2.15\times 10^{10} \Msun/h$, $b=0.68$ & $1.84\times 10^{12} \Msun/h$, $b=0.92$ & $2.10\times 10^{13} \Msun/h$, $b=1.44$ & $2.11\times 10^{15} \Msun/h$\\ 
$z=0.5$ & $6.44\times 10^{10} \Msun/h$, $b=0.84$ & $1.70\times 10^{12} \Msun/h$, $b=1.24$ & $1.53\times 10^{13} \Msun/h$, $b=1.93$ & $1.16\times 10^{15} \Msun/h$\\  
$z=1.0$    & $5.58\times 10^{10} \Msun/h$, $b=1.04$ & $1.44\times 10^{12} \Msun/h$, $b=1.62$ & $1.05\times 10^{13} \Msun/h$, $b=2.50$ & $6.12\times 10^{14} \Msun/h$\\     
\hline
\end{tabular}\\
{\bf Note}. The first row indicates the sample number density, in unit of $(\mpch)^{-3}$.
The minimum halo mass for three number densities and  three redshifts is listed in the table, in units of $\Msun/h$.
In the last column, the maximum halo mass is listed for three redshifts.
\label{tab:haloinfo}
\end{table*}

\begin{figure*}
\epsfxsize=18cm
\epsffile{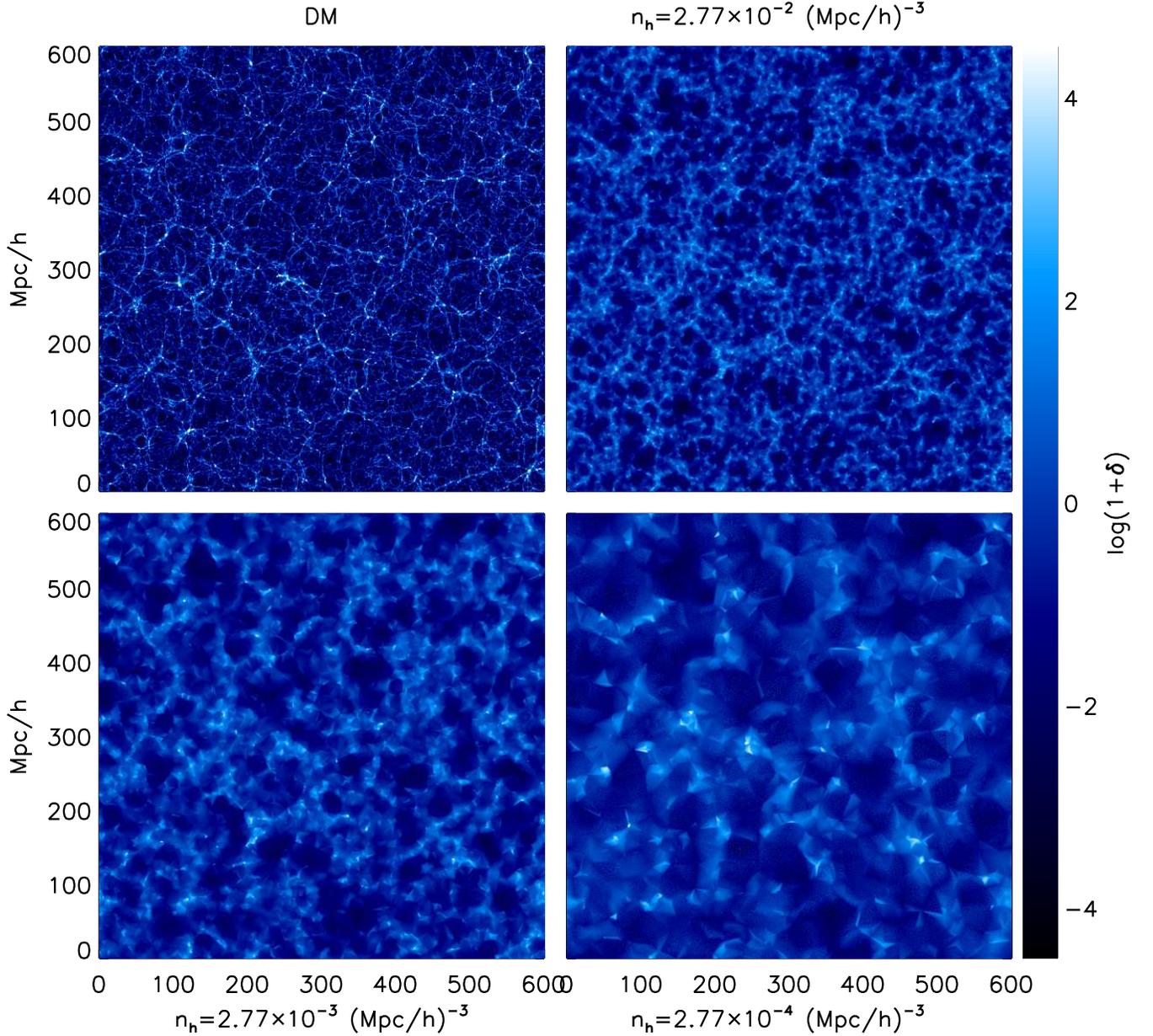}
\caption{Input nonlinear DM density field (top-left) and halo fields with number densities of $2.77\times 10^{-2}$,
$2.77\times 10^{-3}$, and $2.77\times 10^{-4}$ $(\mpch)^{-3}$ for $z=0$.
These halo fields are produced by the DTFE method to avoid empty grid and to improve the stability of the reconstruction algorithm.}
\label{fig:maps}
\end{figure*}

To test the performance of the reconstruction algorithm and to study how far into the nonlinear regime reconstruction works,
we use a simulation run with the \texttt{CUBEP$^3$M} code (see \cite{cubep3m}), involving $2048^3$ dark matter particles in a box with a side length of $600\ \mpch$.
This box size is insufficient for robust direct BAO measurement, which is a signal locate scale of $\sim 100\mpch$.  However, the following results are mainly based on the cross-correlation with the initial conditions.  This box size is sufficient to obtain reliable results due to the cancellation of the sample variance.
The particle-mesh process adopts a $4096^3$ grid and the particle-particle process is involved.
We output the snapshot at $z=0$, $0.5$, and $1$ to cover the typical redshift for current and on-going galaxy surveys.
The reconstruction and analysis are performed on $512^3$ curvilinear and uniform grids.
This grid size ensures that the following analysis is reliable and the reconstruction is not computationally expensive.
We adopt a spherical overdensity halo finder.
The smallest halo contains 10 particles and has a halo mass of $2.15\times 10^{10}\Msun h^{-1}$.
We construct three halo samples with number densities of $2.77\times 10^{-2}$, $2.77\times 10^{-3}$, and $2.77\times 10^{-4}$ $(\mpch)^{-3}$ by setting the lower limit in halo mass.
The detailed halo sample information is listed in Table \ref{tab:haloinfo}.

The halo field is a discrete, highly non-uniform field.
Traditional mass assignment methods (e.g., NGP, CIC, TSC) will leave many grids empty.
As the reconstruction algorithm aims to build a potential isobaric gauge in which the mass per volume element is constant, 
a large area with no matter will cause the algorithm to be unstable.
We need an appropriate mass assignment method to avoid this situation.
We tried both the Voronoi and Delaunay tessellation method (\cite{Bernardeau96,van-de-Weygaert09})
and finally choose to adopt the better performing Delaunay tessellation method by the \texttt{DTFE} code (see \cite{Cautun11}).
It is a linear first-order version of Natural Neighbor Interpolation.
The Delaunay tessellation is constructed from the input halo catalog and
these spatial volume-covering divisions of space into mutually disjunct tetrahedral cells adapt to the local density and the local geometry of the point distribution.
It represents the natural method of reconstructing from a discrete set of samples a volume-covering and continuous density field using the maximum of information contained in the point distribution.
Figure \ref{fig:maps} shows the density slice produced by the DTFE method for simulated dark matter and the corresponding three halo fields at $z=0$.
The triangular shaped color blocks observed in the bottom-right plot is a feature for such a low-density sample.
Next, we apply the reconstruction algorithm on these halo fields to quantify the performance.

\subsection{Number of iteration steps}
\label{sec:iteration}

To maintain positive coordinate transform for each step, a smoothing and limiting scheme is adopted during the reconstruction.
This prevents reaching a final status of exactly constant mass per volume element.
In practice, we use the power spectrum of the reconstructed density field and its cross-correlation coefficient with the initial condition as the convergence criteria.

Comparing results at different iteration steps, we found that we only need $\sim 600$ steps for the reconstruction algorithm to reach convergence for these halo samples, much less than $\sim 1500$ steps for the reconstruction from a nonlinear dark matter density field.
By observing the power spectrum of the reconstructed density field and its cross-correlation coefficient with the initial condition at different iteration steps, we found that the results reach convergence fast on large scales.
For the reconstruction from a nonlinear dark matter density field, most of the computing time is spent on the reconstruction from nonlinear small scales.
This reconstruction method outperforms others by using the information residing in the nonlinear regime (see \cite{zhuhm16c}).
However, the halo field is a discrete sample, in which a part of the small-scale information is missing. 
This comes out as the fast convergence for reconstruction from the halo field.

For a halo sample with lower number density, it contains less usable small-scale information.
The number of iteration steps to reach convergence decreases toward lower halo density.
For convenience, we just fix the number of iteration steps to be 600 for all the halo samples.

\subsection{Dependence on halo number density}
\label{sec:numberdensity}

\begin{figure*}
\epsfxsize=8.5cm
\epsffile{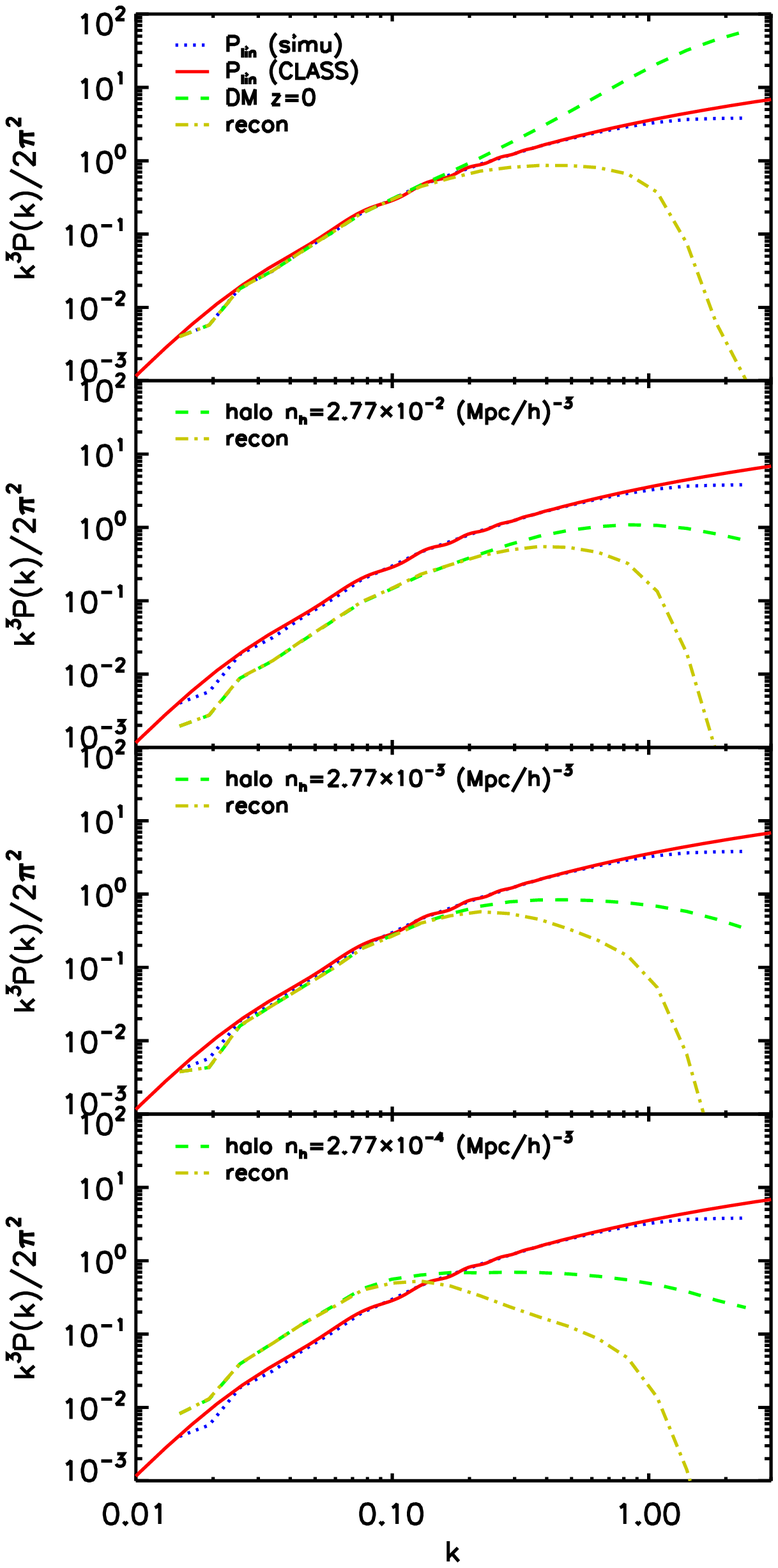}
\epsfxsize=8.5cm
\epsffile{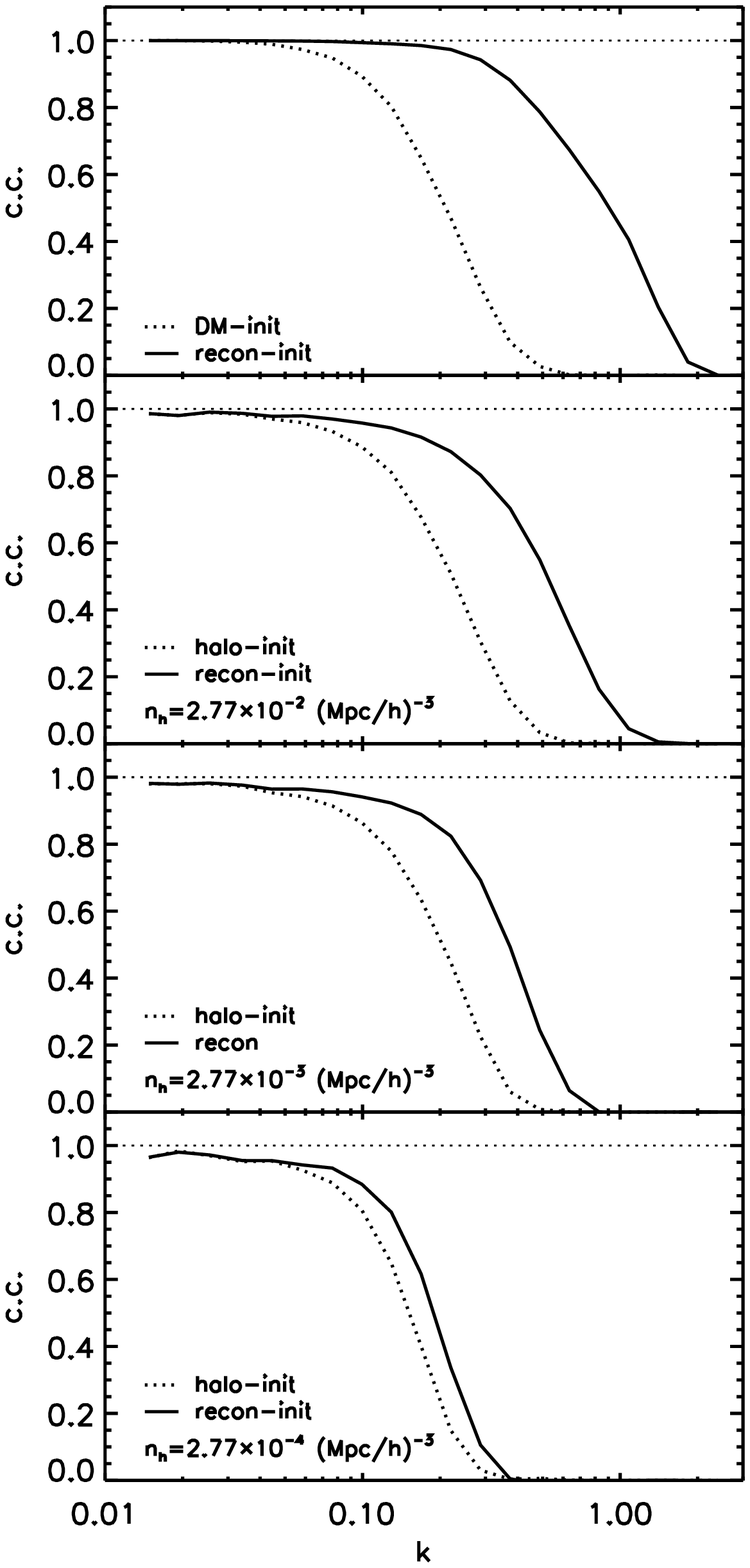}
\caption{Left panel presents the power spectrum of the input nonlinear fields and the reconstructed fields for $z=0$.
As a reference, the linear power spectrum from theory and the simulation's initial condition are also plotted.
Right panel presents the cross-correlation coefficients between the reconstructed fields and the linear density field with the solid lines.
For comparison, the cross-correlation coefficients between the input nonlinear fields and the linear density field are plotted with the dotted lines.
From top to bottom, the input nonlinear field is the DM density field, the halo fields have number densities of $2.77\times 10^{-2}$, $2.77\times 10^{-3}$, and $2.77\times 10^{-4}$ $(\mpch)^{-3}$, respectively.}
\label{fig:powercc}
\end{figure*}

\begin{figure}
\epsfxsize=8.5cm
\epsffile{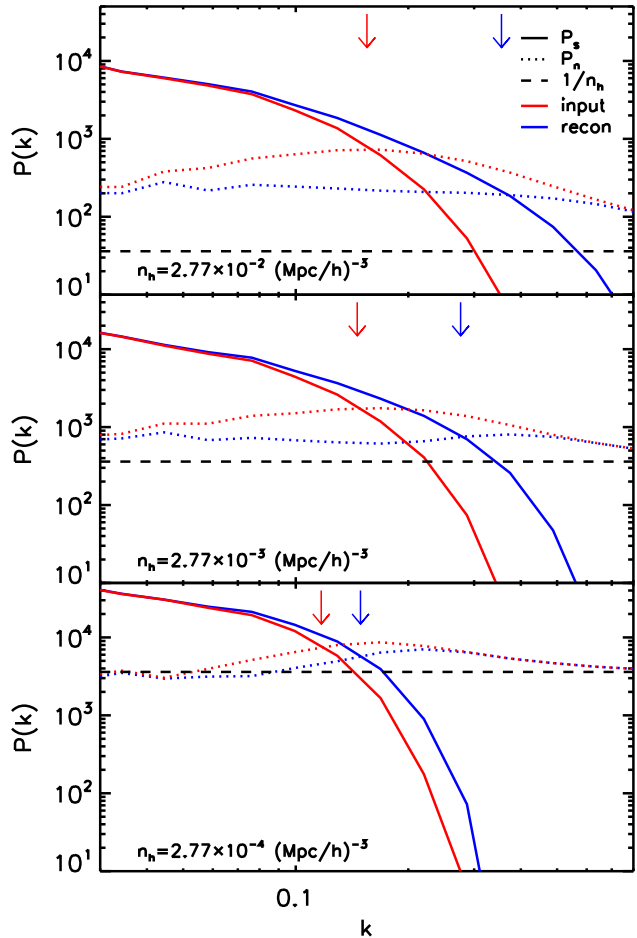}
\caption{Power spectrum decomposition for the input nonlinear (red) and reconstructed (blue) field.
The linear signal term is presented by the solid line, while the noise term is presented by the dotted line.
All the results are scaled mode-by-mode, so that the total power equals to the halo power spectrum with familiar NGP mass assignment.
The expected shot noise for NGP mass assignment, $1/n_h$, is also plotted by the dashed line.
From top to bottom, the input halo field has number densities of $2.77\times 10^{-2}$, $2.77\times 10^{-3}$, and $2.77\times 10^{-4}$ $(\mpch)^{-3}$, respectively.
The downward arrows indicate the scale where the signal term equals the noise term.}
\label{fig:decom}
\end{figure}

\begin{figure}
\epsfxsize=8.5cm
\epsffile{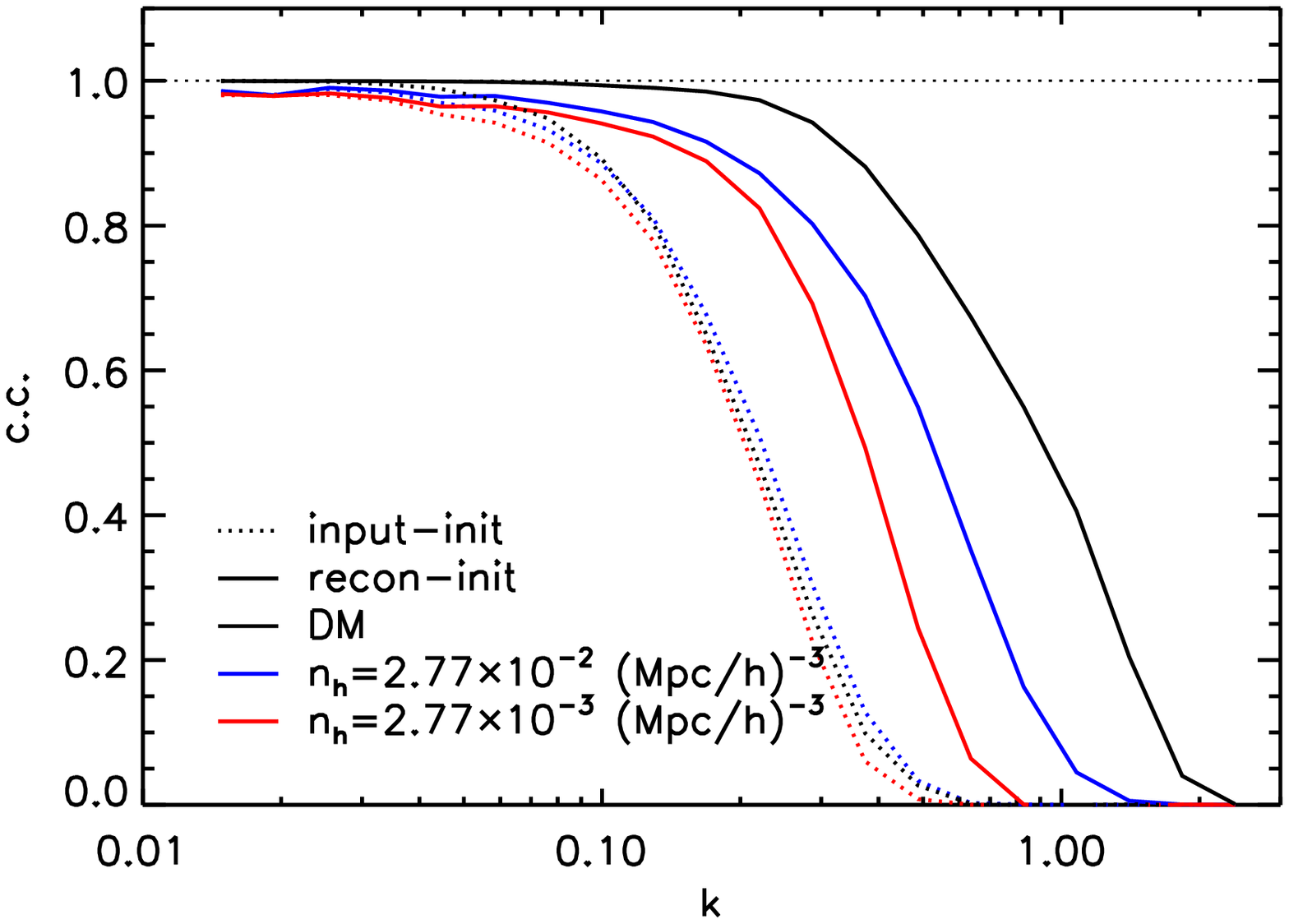}
\caption{Comparison of the cross-correlation coefficients for various input nonlinear and reconstructed fields, including
the DM field and two halo samples.
The three input nonlinear fields have similar coherence with the initial condition.
However, the reconstructed fields benefit from higher number density.}
\label{fig:ccnh}
\end{figure}

In the left panel of Figure \ref{fig:powercc}, we present the power spectrum for the input halo fields and the reconstructed density fields.
The cross-correlation coefficients with the linear density field are presented in the right panel.
As reference, we also plot the result for the reconstruction from the nonlinear dark matter density field at the top row.

The reconstruction process in this work is purely a mathematical approach, 
which does not involve any cosmological dynamics.
Without the assumptions on the galaxy bias, growth rate, and smoothing,
the algorithm always finds the potential isobaric coordinate of which each volume element has approximately constant mass.
This process ensures that the reconstructed density field is the same with the input nonlinear field at a sufficiently large scale.
Halo is a biased tracer of the underlying dark matter density field.
Thus, the reconstructed fields also show the same bias to linear density field at small $k$.
The main purpose of this work is to validate the reconstruction method on the halo field, by investigating the cross-correlation with the initial conditions.
The bias presented in the reconstructed field will not influence the results.
The reconstruction method could be directly performed on a nonlinear density field with the RSD effect.
In this case, the RSD effect is transferred into the reconstructed density field.
From the anisotropy in the reconstructed density field, the RSD effect could be extracted.
We will present the extraction of the RSD information in coming work (H.-M. Zhu et al. 2017, in preparation).

The reconstruction performance is influenced by the number density.
For DM fields, the cross-correlation coefficient curve is shifted toward small scale by a factor of $\sim 6$ by the reconstruction.
We observe a smaller amount of shift (less improvement) toward decreasing halo number density.
For the highest density case, $n_h=2.77\times 10^{-2}$ $(\mpch)^{-3}$ sample, the shift is $\sim 2.5$.
For the intermediate sample, $n_h=2.77\times 10^{-3}$ $(\mpch)^{-3}$, the shift is $\lesssim 2$.
For the lowest density case, only a very small improvement is observed.
Note that we observe quickly damped cross-correlation coefficients for the input nonlinear halo fields toward low number density.
This result is different from the analysis in the literature, e.g. \cite{Mehta11}, which found mild difference in the propagator for $n_h=10^{-3}\sim 10^{-4} (\mpch)^{-3}$ samples.
This implies an additional window function effect induced by the DTFE method.
This could be clearly seen in Fig. \ref{fig:comparestandard} by direct comparison between the cross-correlation coefficient for the input NGP halo field (red dotted line) and the DTFE halo field (blue dotted line).
However, the improvement is still observed by the nonlinear reconstruction for the lowest density case.

Similar to the analysis in \cite{Seo16}, to clearly show how much linear signal is recovered by the reconstruction process,
we decompose the density field into two terms, 
\be
\delta(k)=C(k)\delta_L(k)+n(k)\ ,
\ee
in which $C(k)\delta_L(k)$ is completely correlated with the linear density field.
The pre-factor $C(k)$, often dubbed as the ``propagator'' (e.g., \cite{Crocce06,Crocce08,Matsubara08,Taruya09}), could be obtained by
\be
C(k)=\frac{P_{\delta_L\delta}(k)}{P_{\delta_L}(k)}\ .
\ee
The remaining noise term is just $n(k)=\delta(k)-C(k)\delta_L(k)$.
This decomposition of the density field directly leads to the decomposition of the power spectrum into a linear signal term plus a noise term (also called the ``mode-coupling'' term),
\be
\label{eqn:powerdecom}
P_\delta(k)=P_s(k)+P_n(k)\ ,
\ee
in which $P_s(k)=C^2(k)P_{\delta_L}(k)$ is the linear signal term.
The lower limit in scale where the BAO signal could be robustly measured is quantified by the scale where the linear signal term equals the noise term.

We present the power spectrum decomposition, Equation (\ref{eqn:powerdecom}), for both the input halo fields and the reconstructed fields in Figure \ref{fig:decom}.
Since the DTFE window influences the signal power and the shot noise power on small scales, we correct for such a window function effect before checking whether the noise term is dominated by the shot noise. The correction has no effect on the comparison.
We found that the reconstructed field not only has a larger linear signal term, but also has a lower noise term than the input halo field.
The downward arrows indicate the scale where the linear signal term equals to the noise term.
We clearly see the decrease of the scale (increase in $k$) by the reconstruction process.
The improvement reaches a factor of 2.29, 1.89, and 1.27 for $n_h=2.77\times 10^{-2}$, $2.77\times 10^{-3}$, and $2.77\times 10^{-4}$ $(\mpch)^{-3}$, respectively.
Specifically, for the most concerned case, this scale is decreased from $k=0.16$ to $0.36\ \hmpc$.

Note that the noise term has two contributions, one from the nonlinear clustering and another from the shot noise.
We present the expected shot noise power spectrum for NGP mass assignment in Figure \ref{fig:decom}.
For the highest number density case (top panel), the noise term of the input halo field exceeds the expected shot noise contribution a lot.
This implies that the nonlinear clustering dominates over the shot noise.
After the reconstruction, the linear signal is recovered and meanwhile the nonlinear clustering is suppressed.
Thus, the nonlinear reconstruction is powerful for this case.
For the lowest number density case (bottom panel), the noise term of the input halo field is dominated by the shot noise.
The halo field with such a low number density is almost the combination of the linear signal plus the Poisson noise.
In this case, there is no usable nonlinear information and the reconstruction performance is limited.

In Figure \ref{fig:ccnh}, we present a comparison of the cross-correlation coefficient for various input nonlinear and reconstructed fields, including the DM field and halo samples with $n_h=2.77\times 10^{-2}$ and $2.77\times 10^{-3}$ $(\mpch)^{-3}$.
These three input nonlinear fields have similar coherence with the initial conditions.
This implies that $n_h\gtrsim 2.77\times 10^{-3}$ $(\mpch)^{-3}$ is roughly sufficient for extracting BAO information without any reconstruction process.
However, the reconstruction performance indeed benefits from high halo number density.
This strongly motivates us to go deeper for low-redshift (local) surveys.

\subsection{Performance on the downgraded DM field}
\label{sec:comparedm}

\begin{figure}
\epsfxsize=8.5cm
\epsffile{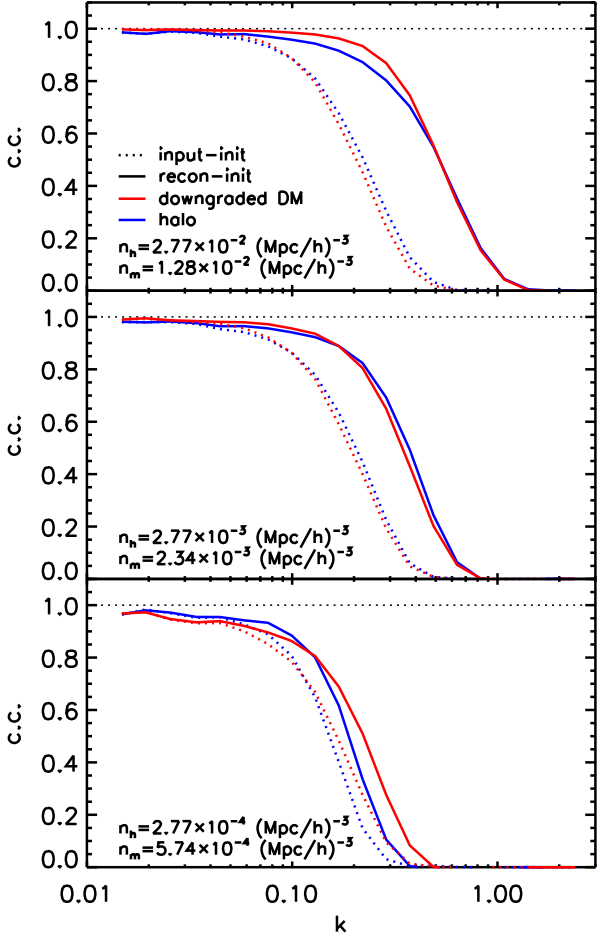}
\caption{Cross-correlation coefficient between the initial density field and the input nonlinear halo/downgraded DM field is presented by the dotted line.  The coefficient between the reconstructed fields and the linear density field is presented by the solid line.
The blue lines are for the nonlinear halo field with decreasing number density from the top to the bottom panel, while the red ones are for the downgraded DM field with the number density $n_m=b^2n_h$.
By construction, these downgraded DM fields suffer from same effective shot noise as the corresponding halo fields.
}
\label{fig:dmdmtot}
\end{figure}

To separate the influence of the shot noise and halo bias on the reconstruction performance,
we construct three downgraded DM samples with the number densities $n_m=b^2 n_h$, in which $n_h$ is the halo number density for a given sample and $b$ is the halo bias.
By this construction, these three downgraded DM samples share the same level of effective shot noise (relative to the amplitude of the power spectrum) as the corresponding halo samples.
We compare the reconstruction performance on the downgraded DM samples and the corresponding nonlinear halo fields
 and the result is presented in Fig. \ref{fig:dmdmtot}.
From top to bottom, the input nonlinear halo/DM fields have decreasing number density.
The dotted lines present the cross-correlation coefficient between the nonlinear fields and the linear density field,
while the solid lines are the results from reconstructed fields.
The red lines are for the downgraded DM samples while the blue ones are for the halo samples.

We found that the improvement by the nonlinear reconstruction is similar for both halo fields and DM fields for each panel.
This similar performance for each case implies that the limiting factor for the nonlinear reconstruction mainly comes from the shot noise of the halo/downgraded DM field.

For the highest number density case, we find that the nonlinear reconstruction has better performance on the downgraded DM sample in the intermediate scales.
This implies that the halo bias also affects the reconstruction performance.
It could be understood from the fact that the estimated displacement also depends on the bias, but in a complicated and nonlinear way.
It is not straightforward to correct it without the modeling of the bias (and the associated dependence on the cosmology).
For example, the linear bias is less than one for this sample.
Correcting this linear bias leads to negative density grids that the nonlinear reconstruction algorithm cannot deal with.
However, the reconstruction performance is expected to be better when we find methods to properly deal with the halo bias first.
We leave this for future investigation.

For the number density of the $2.77\times 10^{-3} (\mpch)^{-3}$ case, the halo sample roughly has $b\sim 1$, and shares the same shot noise with the downgraded DM sample.
Thus, it is expected that the reconstruction has the same performance in these two samples. 

For the bottom panel, the number density is low, and thus the DTFE window function is heavy and sensitive to both the number density and the clustering property of the sample.
We observe different cross-correlation coefficients for the nonlinear halo and DM sample.
Thus, the direct performance comparison is less meaningful.
However, we observe a similar amount of increase in the cross-correlation coefficient by the nonlinear reconstruction process for both samples.

\subsection{Comparison with the standard reconstruction}
\label{sec:compareeisen}

\begin{figure}
\epsfxsize=8.5cm
\epsffile{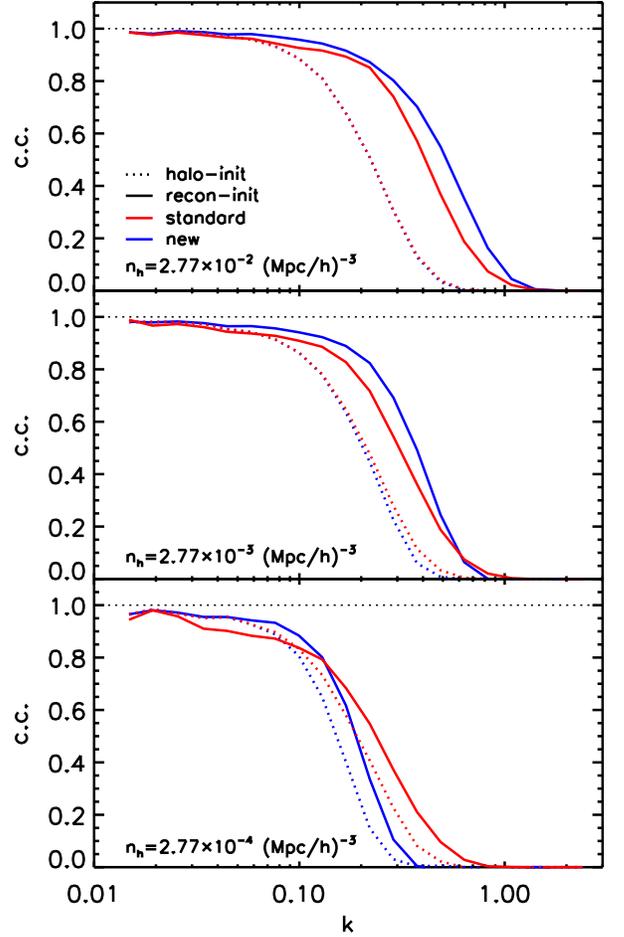}
\caption{Cross-correlation coefficient between the reconstructed fields and the linear density field is presented by the solid line.
For comparison, the result for the input halo fields is presented by the dotted line.
The red lines are for the standard reconstruction method, while the blue ones are for the nonlinear reconstruction.
From top to bottom, the input nonlinear halo fields have decreasing number density. }
\label{fig:comparestandard}
\end{figure}

We compare the reconstruction performance with the standard method \cite{Eisenstein07} used in the literature for $z=0$.
The performance of standard reconstruction depends on the smoothing kernel used in the algorithm.
Here, we just choose to use a Gaussian smoothing with a smoothing length of  $15 \mpch$.
We quantify the performance by the cross-correlation coefficient with the initial conditions and 
the result is presented in Fig. \ref{fig:comparestandard}.
The blue solid lines are the cross-correlation coefficient from the nonlinear reconstruction method,
 while the red solid lines are from the standard reconstruction method.
The dotted lines represent the cross-correlation coefficient between the input nonlinear halo fields and the initial conditions.
From top to bottom, the input nonlinear halo fields have decreasing number density. 
We find that for the two high number density case, the nonlinear reconstruction method outperforms the standard one.
Note that this outperformance is not as significant as in the DM case, due to the limitation from the shot noise in the halo fields.

For the lowest number density case, both the nonlinear reconstruction method and the standard method increase the cross-correlation with the initial conditions (from dotted line to solid line).
However, the nonlinear reconstruction method outperforms the standard one at $k\lesssim 0.13 \hmpc$ and vice verse at $k \gtrsim 0.13 \hmpc$.
Note that the cross-correlation coefficient of the input nonlinear halo field for the standard method (red dotted line) is calculated from the halo field with NGP mass assignment, which has a better correlation with the initial conditions than the DTFE method (dotted blue line).
This implies that for such a low number density case, the DTFE method loses part of the linear information due to the heavy window function effect.
This results in the worse performance of the nonlinear reconstruction method at $k\gtrsim 0.13 \hmpc$.

\subsection{Dependence on redshift}
\label{sec:evo}

\begin{figure}
\epsfxsize=8.5cm
\epsffile{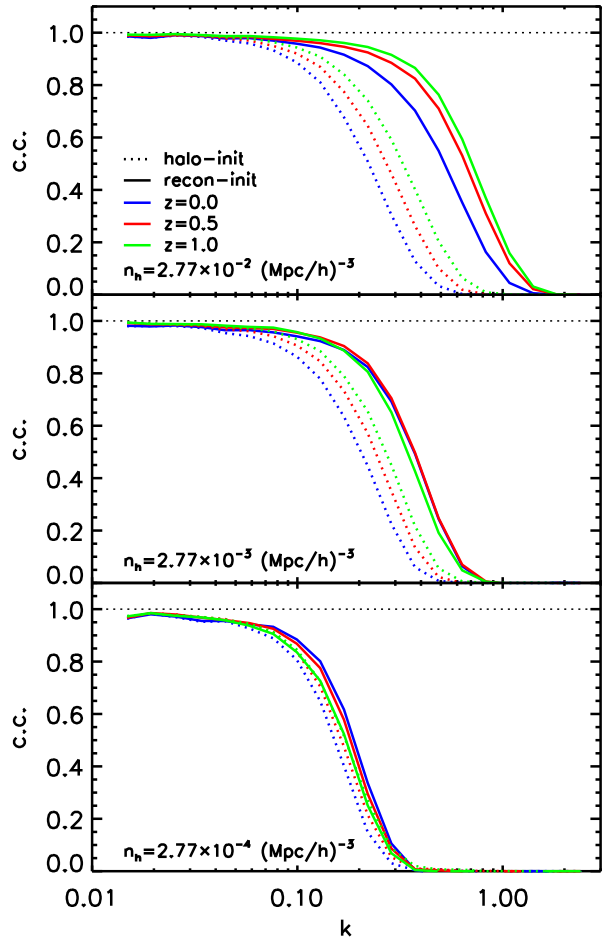}
\caption{Cross-correlation coefficient between the reconstructed fields and the linear density field is presented by the solid line.
For comparison the result for the input halo fields is presented in dotted line.
For a given halo number density, the result for $z=0$, $0.5$, and $1$ is presented in blue, red, and green, respectively.}
\label{fig:numevo}
\end{figure}

We present the redshift dependence of the reconstruction performance for each halo sample in Figure \ref{fig:numevo}.
Note that these halo samples at different redshifts are not the same halo population at different cosmic epochs, since we just choose a specific halo number density for all three redshifts.

The top panel shows the result for the input halo sample with a number density of $2.77\times 10^{-2}$ $(\mpch)^{-3}$. 
The higher redshift halo field experiences less nonlinear evolution, and thus contains more linear information.
The reconstruction increases the cross-correlation coefficients in a similar way for all of the three redshifts we investigated.
The performance is only mildly degraded for $z=1$.

For the input halo field with a number density of $2.77\times 10^{-3}$ $(\mpch)^{-3}$, we can still see the improvements for all  three of the redshifts.
However, the improvement is less obvious toward higher redshift.
This could be explained by the fact that the higher redshift halo field has not only a lower nonlinear clustering effect, but also a relatively larger shot noise contribution.
Thus, the power of the nonlinear reconstruction is limited.
As a result, the reconstruction recovers the cross-correlation to a comparable level for all three of the redshifts.

For the case in which the number density is $2.77\times 10^{-4}$ $(\mpch)^{-3}$, the input halo fields at three redshifts contain similar coherence with the initial condition.
Meanwhile, their noise term is dominated by the large Poisson noise.
After the reconstruction, we only observe very small or no improvement for these redshifts.

In the two lowest density cases, we observe the convergence for the recovered cross-correlation coefficients for different redshifts.
If we define a characteristic scale to quantify the performance, namely where the cross-correlation coefficients drops to 0.7 (corresponds to $S/N=1$ for the power spectrum decomposition Equation (\ref{eqn:powerdecom}), see \cite{zhuhm16a}),   
we find that this roughly scales as $n_h^{1/3}$.
The fact that the dependence of this characteristic scale on $n_h$ with a power of $1/3$ rather than a power of $\ll 1/3$ implies that the reconstruction performance is already limited by Poisson noise.
It is also this limitation that mildly degrades the reconstruction performance for $n_h=2.77\times 10^{-2}\ (\mpch)^{-3}$ at $z=1$.


\section{Conclusion and Discussion}
\label{sec:conclusion}

We tested the reconstruction method proposed in \citet{zhuhm16c} on the simulated halo fields with three different number densities and at three redshifts.
The reconstruction performance is quantified by the extension in scale where the BAO signal could be robustly measured.
We decomposed the power spectrum of both the input halo field and the reconstructed field into a linear signal term plus a noise term.
This scale is defined at where the linear term equals the noise term.
For the most concerned case, $n_h=2.77\times 10^{-2}\ (\mpch)^{-3}$ at $z=0$, which is close to the condition of the SDSS main galaxy sample, we found that the improvement reaches a factor of $2.29$ in scale (from $k=0.16$ to $0.36\ \hmpc$), or equivalently, a factor of $12$ for available modes for BAO measurement.
For this case, the nonlinear reconstruction method outperforms the standard reconstruction method.

The reconstruction performance depends on the halo number density.
At $z=0$, the improvements reach a factor of $2.29$, $1.69$, and $1.15$ for the halo sample with a number density of $n_h=2.77\times 10^{-2}$, $2.77\times 10^{-3}$, and $2.77\times 10^{-4}$ $(\mpch)^{-3}$, respectively.
We also reported the limited reconstruction power for the low number density cases, which roughly correspond to the targets of the on-going high-redshift surveys.
By testing the reconstruction performance on properly constructed downgraded DM samples, we found that the main limiting factor for the reconstruction performance is the heavy shot noise for low-density samples.

To apply the new reconstruction method on discrete halo fields,
we need a suitable mass assignment method, which does not induce an extra degree of freedom in the process.
The DTFE method we adopted in this work is a straightforward attempt.
We also tried other methods like the Voronoi tessellation.
The performance is a little bit downgraded by adopting the Voronoi tessellation method.
There may exist a better mass assignment worth trying in dealing with this issue.
However, for the most concerned high-density case, the choice of mass assignment is less important.
For the low-density case, none of the mass assignment could present the underlying DM field well without introducing extra parameters.
We also reported that the heavy window function effect from the DTFE method erases a part of the linear signal in the nonlinear halo field for the lowest number density case.

As discussed in \cite{zhuhm16c},
the Lagrangian BAO reconstruction algorithm involves displacing individual objects according to the linear displacement that is computed from the observed galaxy distribution under certain model assumptions (the smoothing scale, galaxy bias, growth rate, etc.; see \cite{Eisenstein07}). The results depend on the assumed fiducial model and must be tested against different parameter choices, which is computationally expensive (see \cite{Padmanabhan12}). The reconstruction method used in this work is a purely mathematical approach with no cosmological dynamics involved. 

The reconstruction from the nonlinear DM field gives an estimate for the displacement field (See \cite{zhuhm16c}).
This is not the case for the reconstruction from a biased tracer, i.e. the halo field.
The reconstructed displacement from a biased halo sample does not respond to the halo displacement in the Lagrangian halo formation scenario.
We also identified the influence of the halo bias by performing reconstruction on the downgraded DM samples with the same number density as the corresponding halo samples.}
One could involve the above mentioned model assumptions (the fiducial cosmology, galaxy bias, growth rate, etc.) to correct the bias and RSD effect prior to the reconstruction process.
In this way, the reconstructed displacement field is more physically motivated and might have further applications.
We leave this to future investigation.

We focused on investigating the dependence on the halo number density and redshift.
One immediate and urgent future work is the performance test on the halo samples with the RSD effect.
The RSD effect is due to the structure growth, and thus contains important cosmological information.
The observed position of an object is shifted by its peculiar velocity along the line of sight.
This simply adds an extra offset on the real displacement field. 
Thus, the displacement field reconstructed from the observed density field automatically includes this additive offset. 
Since much fewer nonlinearities are involved in the displacement potential, the measurement and modeling of RSD will be improved significantly.
This also helps to simultaneously model the BAO and RSD signal.

To apply the proposed reconstruction method in observations, there are many observational issues that require investigation, including the selection function, the survey geometry, and etc.
We will investigate these issues in the near future.

\section*{Acknowledgement}
We thank the anonymous referee for the useful suggestions that made the article more informative.
We thank Pengjie Zhang, Xin Wang, Xuelei Chen, Kwan Chuen Chan, Haoran Yu, Morgan Bennett, and Qiaoyin Pan for useful discussions.
This work was supported by the National Science Foundation of China (grant No. 11403071).
The simulation was performed on the BGQ super-computer at the SciNet HPC Consortium. SciNet is funded by the Canada Foundation for Innovation under the auspices of Compute Canada, the Government of Ontario, the Ontario Research Fund Research Excellence, and the University of Toronto. 
The Dunlap Institute is funded through an endowment established by the David Dunlap family and the University of Toronto. Research at the Perimeter Institute is supported by the Government of Canada through Industry Canada and by the Province of Ontario through the Ministry of Research \& Innovation.

\bibliographystyle{aasjournal}
\bibliography{yuyu}

\clearpage

\end{document}